# Determinants of Longevity amongst SSNIT Pensioners in Ghana.

Bernard Gyamfi*

*Postgraduate Student, Department of Statistics and Actuarial Science, College of Science, Kwame Nkrumah University of Science and Technology, Kumasi, Ghana*

This paper statistically analysed pensioner longevity in Ghana. It fundamentally sought to ascertain the significant determinants of longevity amongst pensioners in Ghana, more specifically SSNIT pensioners by estimating the mortality rate of SSNIT pensioners in Ghana, determining the factors that significantly affect the longevity of the average Pensioner in Ghana and constructing a predictive model for predicting the longevity of SSNIT pensioners in Ghana. Secondary data was obtained from the leading pension provider in Ghana, the Social Security and National Insurance Trust, which is mandated to manage the First-tier pension scheme in Ghana. The paper employed regression – based models at different sections. Parameter testing and model diagnosis were rigorously conducted to ascertain the significant determinants of longevity. The results of the study revealed that the total number of male deaths was significantly greater, about four times more, than the total number of female deaths. There was sufficient evidence that there exists a lower rate of death among female pensioners as compared to male pensioners. Furthermore, a significant number of the SSNIT pensioners used in the analysis survived less than 8 years after retirement before their death and the average basic salary of the pensioners who lived less than 8 years after retirement was GH₵ 7741.827 and served for 35.65 years on average. On the contrary, it was observed that pensioners who lived more than 8 years after retirement had an average basic salary of GH₵ 5544.20 and served for 31.49 years on average. In conclusion, predictors such as basic salary, the number of years of service in the workforce, the age of the pensioner before death and the gender of the pensioner were statistically significant in predicting the number of years a SSNIT pensioner survived after retirement before death that is, whether a particular SSNIT pensioner survived less than 8 years or 8 years and above after retirement. The authors recommend to Ghanaian employees that, their health should be of a great priority as the number of years of service in the workforce keenly determines the number of years they would survive after retirement before death as every increase in the service years increases the likelihood of a SSNIT pensioner surviving less than 8 years after retirement by 22.36%.

## I. INTRODUCTION

Human mortality has been on a decelerating trajectory around the world since the beginning of the twentieth century (Wrigley-Field, 2019; Wong-Fupuy & Habberman, 2013). There is no philosophical skepticism as to the assertion that people live longer today than some years back in time. This can be attributed to a lot of factors, some of them being the medical and technological breakthroughs in this century. Understanding human mortality trends is vital to many researchers, governments, demographers, actuaries, and policymakers. Due to this, research in this field progresses very importantly and has gained immense recognition, since these mortality trends are needed for policymaking, government expenditures, pension, and insurance, to mention but a few. Most developing countries have less knowledge of the human mortality trends and longevity of their citizens since less energy has been channelled to these fields, however, developed countries focus more on these fields and have been well represented in the areas of human mortality trends and longevity. Most of the developing countries within the sub-Saharan region massively and excessively rely on international institutions and organization such as the UN and the World Health Organization for their human mortality data (Ofori-Amanfo, 2019; OECD, 2022).

---

*Corresponding author's e-mail: bgyamfi43@st.knust.edu.gh/bernard.gyamfi01@gmail.com.

Ghana's life expectancy as of 2021 was 64.42 years, which represented an increase of 0.4% from 2020, which was 64.17 years as compared to 63.19 and 63.65 years in 2019 and 2018 respectively (Macrotrends, 2021). As a result, it is an indisputable fact that Ghana is one of the countries in Sub-Sahara Africa that is experiencing rapid growth in its elderly population (Kpessa-Whyte & Tsepko, 2020). Since old age comes with a lot of health complications and implications, the government and policymakers introduced the Social Security system. Social security is the protection that society provides to individuals and households to ensure access to health care and to guarantee income security, particularly in cases of old age, unemployment, sickness, invalidity, work injury, maternity, or loss of a breadwinner (Kumado & Gockel, 2003). This in effect yielded the idea of pension. Pension, in simple terms, is a fund into which a sum of money is paid during an employee's employment years and from which payments are drawn to support the person's retirement from work in the form of periodic payments or life annuities. In Ghana, the regulatory body which oversees pensions is the National Pensions Regulatory Authority (NPRA). There are a lot of these social security systems institutionalized by governments around the world, some of them are the Social Security Administration in the United States and The Department of Works and Pensions (DWP) in the United Kingdom. In Ghana, the Social Security and National Insurance Trust (SSNIT) manages and coordinates social security established in the statutory of the National Pensions Act 766, 2008.

The Social Security and National Insurance Trust (SSNIT) is the major leading pension provider in Ghana. It is a statutory public trust charged, with the administration of Ghana's Basic National Social Security Scheme. Its mandate is to cater for the mandatory First Tier of the Three-Tier Pension Scheme introduced under the National Pensions Act, Act 766 of 2008 replacing the CAP 30 due to concerns and protests concerning the level of pensions' inadequacy to support a respectable lifestyle for retired public personnel were made by employees in 2004 which led to the commencement of a significant reform of the pension system in July 2004 by the government of Ghana in response to workers concern about the need to guarantee a universal pension scheme for all employees in the nation. A Presidential Commission on Pensions was established with a focus on the public sector, the commission was tasked with reviewing current pension plans and formulating sound proposals for feasible pension plans that would provide retirement income security for Ghanaian workers. In March 2006, the Commission turned in its final report (Acheampong, 2021; SSNIT, 2022). The new contributory three-tier pension plan consisted of a voluntary plan and two mandatory plans, i.e., a first-tier, compulsory, basic national social security program that will include an enhanced SSNIT benefit system, a second-tier occupational (or work-based) pension plan that is privately administered, mandated for all employees, and principally created to provide subscribers with better lump sum benefits and to offer additional money for employees who want to make voluntary contributions to improve their pension benefits as well as for employees in the informal sector, there should be a third-tier voluntary provident fund and personal pension schemes supported by tax benefit incentives subjected to the following requirement for qualification; an individual must be at least 60 years old and have contributed for a minimum of 180 months (15 years) under Act 766 and 240 months (20 years) under PNDCL 247 to be eligible for a full pension (NPRA, 2022). Currently, the Pension Scheme as administered by SSNIT has an active membership of over 1.6 million as of January 2021 with over 226,000 pensioners who regularly receive their monthly pensions from the scheme (SSNIT, 2022).

Longevity risk can be defined as the chance that life expectancies and the actual survival rates exceed expectations or pricing assumptions, resulting in greater-than-anticipated cash flow needs on the part of insurance companies or pension funds (AON, 2022) In simple terms

Longevity risk is, therefore, the risk associated with pension funds or insurance companies when assumptions about life expectancies and mortality rates are not veracious (Sinclair, More, & Franklin, 2014). Mortality on the contrary is the number of deaths in a population during a given time. Mortality rate, or death rate, is a measure of the number of deaths (in general, or due to a specific cause) in a particular population, scaled to the size of that population, per unit of time (Ofori-Amanfo, 2019). According to the American Center for Insurance Policy and Research of the National Association of Insurance Commissioners (NAIC), key drivers of the growing need to address longevity risk include an ageing population, increasing life expectancy, a shift in who bears the responsibility of sufficient retirement income, uncertainty of government benefits and economic volatility (Blake & Turner, 2013). World trends suggest that social instability is noticeable in all developing countries pursuing structural adjustment policies and all societies in the transition generally. Indeed, the world trend has shown that there is considerable concern about social protection, especially pension and retirement issues, even in developed countries. It is generally known that old age security systems are declining in birth rates and increased longevity (Kumado & Gockel, 2003).

With a knowledge gap as to whether or not, factors such as gender of pensioner, number of years lived after retirement, number of working years, replacement ratio, and monthly pension benefits, either from the Defined Benefit (DB) or the Defined Contribution (DC) do have a significant impact on the number of years a pensioner would be alive after attaining the retirement age of 60 years for the case of Ghana and 65 years in the United States and other parts of the world (Madrigal, Matthews, Patel, Gaches, & Baxter, 2011). In summary, every stakeholder faces longevity risk, from Governments to private DB schemes and national defined retirement benefits providers and other lifetime annuity providers. However, when longevity is "uncovered", the individual bears the full extent of the longevity risk.

As to this fact, this paper seeks to determine the significant determinants of longevity amongst pensioners in Ghana, more specifically SSNIT pensioners by estimating the mortality rate of SSNIT pensioners in Ghana, determining the factors that significantly affect the longevity of the average Pensioner in Ghana and constructing the most accurate predictive model for predicting the longevity rate among SSNIT pensioners in Ghana.

This paper would be of immense assistance to policymakers, insurance companies and the government in planning policies by determining which factors significantly contribute to the mortality and longevity of SSNIT Pensioners in Ghana. Again, it would assist pensioners and workers in planning their pension plans. Workers would be able to determine the impact of longevity risk on their lives as retirees and prepare adequately for a good standard of living during retirement since they would be exposed to the vital factors they are to be observe, which greatly contribute to the mortality and longevity of pensioners. Furthermore, the paper would be relevant to academicians by contributing to the existing body of knowledge in the fields of mortality and longevity of SSNIT pensioners in Ghana.

The subsequent sections of this paper are structured into four sections. The first section of this paper explores and discusses the various existing literature relating to the topic. In Section 2, we describe the various methodologies and the model fitting, selection and prediction accuracy of the models utilized in this paper. Furthermore, the fourth section presents the findings of paper and a thorough discussion of the findings evaluating their applicability.

## II. REVIEW OF RELEVANT LITERATURE

Antolin (2007) investigated and studied how the liabilities of employer-provided defined-benefit (DB) private pension plans were impacted by uncertainty surrounding future mortality and life expectancy outcomes, or longevity risk. The study stated that a stochastic technique to estimate mortality and life expectancy was superior because it allowed probabilities to be attached to various forecasts, which was necessary to assess uncertainty and related hazards effectively. Regarding this, their study offered the findings of the Lee-Carter model estimation for several nations that are members of the Organization for Economic Co-operation and Development (OECD). Moreover, it used the Monte-Carlo simulations of the Lee-Carter model to illustrate the uncertainty around future mortality and life expectancy results. To evaluate the impact of longevity risk on defined-benefit pension plans employers offer. He looked at the various methods that private pension plans used to account for longevity risk in actuarial computations. Unfortunately, the majority of pension plans failed to fully factor in anticipated increases in life expectancy and death. Estimates of the range of increases in the net present value of annuity payments for a hypothetical defined benefit pension fund were provided by the study Lastly, the study covered several policy concerns with longevity risk, highlighting the necessity of a unified strategy. Nevertheless, the research was restricted to defined benefit (DB) private pension plans offered by employers, ignoring alternative strategies like defined contribution (DC) pension plan providers, and it was conducted solely in OECD nations, excluding developing nations.

However, Ofori-Amanfo (2019) pointed out that in the recent past, hedging against longevity risk and projecting mortality trends have garnered a lot of attention during a time when life expectancies are unexpectedly rising. As a result of improvements in hygienic practices, technology, and medicine, life insurance companies may benefit financially from lower liabilities resulting from lower death benefit payouts. However, annuity insurance plans like pensions may experience losses as a result of longer lifespans.

The research examined whether longevity risk exists in Ghana and what financial strategies SSNIT should implement for retirees who live longer than the 15-year guaranteed period. Lee-Carter and Cairns-Blake-Dowd models were used to predict mortality patterns using secondary source data from SSNIT that covered the years 1991 to 2017. The investigation's findings demonstrated that, although life expectancy was increasing, death rates were declining. The study also called for prudent financial management and operational efficiency to ensure long-term solvency, noting that since longevity was improving, SSNIT could hedge individual longevity risk by obtaining a life annuity from life insurance companies.

However, his study failed to assess the determinants of these longevity risks amongst the SSNIT pensioners in Ghana. As to that fact, this paper determines the factors that significantly affect the longevity of the average Pensioner in Ghana and further builds a predictive model for predicting the longevity risk among SSNIT pensioners in Ghana using regression modelling.

Similarly, Adeyele & Igbinosa, (2015) used Nigeria as a case study to simulate the sufficiency of retirement pay for university personnel. They suggested that to provide sufficient retirement income for existing university employees, they must comprehend how income changes after retirement. If not, they run the risk of losing their mental and physical health once they retire, particularly if they have no other source of income but their pension. Their study looked at the minimum contribution rates that workers under defined contribution (DC) plans must pay to save enough money throughout the DC's accumulation period to live comfortably after retirement. The study simulated replacement ratios for all categories of university employees using various parameters, including entry age, year of service, starting salary, Consumer Protection Inflation (CPI), and contribution rates. This helped employers and employees assess how much should be contributed each month to cover the costs of the employees' retirement. The study found that early retirement may be harmful to employees' accumulated savings at the payout phase of retirement and that both the present 18% contribution rate and the 15% monthly contribution rate between June 2004 and July 2014 are egregiously inadequate. The study also suggested, among other things, that to guarantee the sufficiency of pension income for retiring university personnel in Nigeria, the contribution rates should be reviewed upward per the nation's economic circumstances. The authors did not, however, attempt to determine how the longevity risk of university personnel and retirement income interacted.

Cairns (2013) conducted a study of the current status of using stochastic models to measure and manage longevity risk. His discussion centred on the models' resilience to diverse inputs, an aspect that holds significant importance when devising a risk management plan. Robust multi-populational mortality models still required a great deal of work in the modelling domain, and in the risk management domain, we needed to gain a deeper comprehension of the goals of pension plans and how they should be maximized.

In the end, the study offered several recommendations for further work on both fronts. On the contrary, this paper investigates the modelling of longevity among SSNIT pensioners in Ghana using linear regression models, as opposed to stochastic methods. (Booth & Tickle, 2008) noted that although regression-based (GLM) approaches have been less successful because of time-related nonlinearities, ongoing advances in life expectancy beyond previously accepted limitations have highlighted the vital significance of mortality forecasting. Their research examined key advancements in mortality forecasting from 1980 to the present using three main categories: extrapolation, explanation, and expectancy.

Because expectations are subjective and expert expectations are almost always conservative, there is not an adequate basis for mortality forecasting. However, explanation is limited to specific causes of death that have established causes. Decomposition by cause of death presented by issues with data difficulties and reasons that are not independent of one another. Statistical techniques were employed in most advancements, which involved extrapolative forecasting instead of age-specific graduating models. The most effective approaches, though, have been those that employed two-factor models (age-period or age-cohort). The estimation of forecast uncertainty has been enhanced by recent developments, and the two-factor Lee-Carter technique, and especially its variations, has proven successful in terms of accuracy. More recent approaches include three factors; the Lee-Carter age-period-cohort model seems to be a promising one. There is now available specialized software that has been developed. The study also made clear the need for more comparative analyses of approaches concerning the precision and uncertainty of the point forecast, covering a broad variety of death scenarios.

In contrast to the stochastic models that (Lee & Carter, 1992) and (Cairns, 2013) suggested be used for the measurement and management of longevity risk, the Lee-Carter Models are appropriate for actuarial applications for several reasons, including the model easily interprets a relatively small number of parameters. Remarkably, (Cairns, Blake, & Dowd, 2008) suggested that certain characteristics should be present in a decent model in order to determine its suitability. Among the attributes are the following: Invariance with past data, Implementation simplicity, Parsimony, Transparency, Sample path and forecast intervals, and Uncertainty. Therefore, this paper explores the Regression-based models in determining the significant determinants of longevity amongst SSNIT pensioners in Ghana.

## III. MATERIALS AND METHOD

### A. Description of Dataset

Secondary data was obtained from the Social Security and National Insurance Trust (SSNIT) which is the leading pension provider in Ghana across all sectors of the Ghanaian economy with about 2.7 million total contributors as at 2018 (NPRA, 2019). The data represents Ghanaian SSNIT pensioners in the SSNIT pension scheme from 2015 to 2020. Since one of the sources of data used in longevity modelling is from pension plans and insurance companies, the data obtained included the date of birth, entry dates, date of retirement, gender, date of death, monthly pension(gratuity) and mortality rate of both male and female SSNIT pensioners making a total of 12,868 data points. It should be noted by the reader that this data set was prior to the outbreak of the pandemic in Ghana.

Furthermore, less emphasis was placed on early retirement, as to that fact all analysis was based on the normal mandatory pensionable age of 60 in Ghana. i.e., old age pension. However, others such as invalidity pension, survivor's lump sum pensions, reduced pension and earned pension right were not considered in this paper.

### B. Model Description

#### 1. Mortality Rates

Age-Specific Death Rates

The age-specific death rate is the total number of deaths of people of a specified age or age group in a specified geographical area divided by the population of the same age or age group in the same geographical area and multiplied by 1,000.

$$ASSDR = \frac{Total\ Deaths\ in\ specified\ age\ group}{Total\ number\ of\ persons\ in\ that\ age\ group} \times 1,000$$

(1)

or these are ratios of deaths by age of persons in each age interval, usually in 5-year age interval.

$$ASDR_X = \frac{D_X}{P_X} \times 1000 \qquad (2)$$

where

$D_X$: is the number of deaths during the year in the age interval x

$P_X$: is the mid-year population in the same age group.

### Sex-Specific Death Rates

The sex-specific death rate is the total number of deaths of people of a specified sex group in a specified geographical area divided by the population of the same sex group in the same geographical area and multiplied by 10,000.

$$SSDR = \frac{Total\ Deaths\ in\ specified\ sex\ group}{Total\ number\ of\ persons\ in\ that\ sex\ group} \times 1,000 \qquad (3)$$

## 2. Regression - Based Models

### Binary Logistic Regression

The binary logistic regression model shows how the outcome variable which is a binary categorical variable depends on the sets of explanatory variables(predictors). It comprises of the distribution of Y is a Binomial, the link function (logit) and the systematic components

$$\eta = logit(\pi) = \ln\left(\frac{\pi}{1-\pi}\right) = \beta_0 + \beta_1 x_1 + \beta_2 x_2 + \cdots + \beta_k x_k = x^T_i \beta \qquad (4)$$

Where, $E(Y) = \pi$, $logit(\pi)$ is the link function and $x^T_i \beta$ is the systematic component which are the explanatory variables (continuous, discrete or both) and $\beta$ are parameters to be estimated.

The general form of the multiple logistic regression model is

$$\pi = \frac{e^{\beta_0 + \beta_1 x_1 + \beta_2 x_2 + \cdots + \beta_k x_k}}{1 + e^{\beta_0 + \beta_1 x_1 + \beta_2 x_2 + \cdots + \beta_k x_k}} \qquad (5)$$

Where $p$ is the number of independent variables and $P(Y = 1|x) = \pi$ is the conditional probability that an outcome is present.

Specifying the multiple logistic regression model to this study, the logit of the multiple logistic regression model is given by;

$$g(x) = \beta_0 + \beta_1 + \beta_2 + \beta_3 + \beta_4 \qquad (6)$$

In this case, the logistic regression model is

$$P(Y = 1|x) = \pi(x) = \frac{e^{g(x)}}{1+e^{g(x)}} \qquad (7)$$

$$\pi = \frac{e^{\beta_0 + \beta_1 \lambda_1 + \beta_2 \lambda_2 + \beta_3 \lambda_3 + \beta_4 \lambda_4}}{1 + e^{\beta_0 + \beta_1 \lambda_1 + \beta_2 \lambda_2 + \beta_3 \lambda_3 + \beta_4 \lambda_4}} \qquad (8)$$

where, $\lambda_1$ = basic salary of pensioner, $\lambda_2$ = the number of years of service in the workforce, $\lambda_3$ = the gender of the pensioner, $\lambda_4$ = age of pensioner before death. Again, longevity is categorized as

$$y_i = \begin{cases} 1 & \text{if the pensioner had survived less than 8 years after retirement} \\ 0 & \text{if the pensioner had survived 8 or more years after retirement} \end{cases}$$

### Odds Ratio

Generally, the odds of the outcome being present among pensioners with Y = 1 defined as

$$\pi(1) = \frac{\pi(1)}{1-\pi(1)} \qquad (9)$$

Similarly, the odds of the outcome being present among pensioners with Y = 0 defined as

$$\pi(0) = \frac{\pi(0)}{1-\pi(0)} \qquad (10)$$

With regard, the odds ratio (OR) is defined as the ratio of the odds for Y = 1 to the odds for Y = 0, is given as;

$$\pi(1) = \frac{\pi(1)/1-\pi(1)}{\pi(0)/1-\pi(0)} \qquad (11)$$

Hence, for logistic regression with a dichotomous independent variable coded 1 and 0, the relationship between the odds ratio is the regression coefficient given as

$$OR = e^{\beta_1} \qquad (12)$$

### Wald Test

The Wald test is obtained by comparing the maximum likelihood estimate of the slope parameter $\hat{\beta}_1$ to an estimate of its standard error. The resulting ratio will follow a standard normal distribution. This is used in testing for the significance of a variable. Mathematically;

$$W = \frac{\hat{\beta}_1}{SE(\hat{\beta}_1)} \quad (13)$$

### 3. Assessing the logistic model

Goodness-of-fit is mostly used in describing how effective our model is. It is assessed over the constellation-fitted values determined by the covariates in the model, not the total collection of covariates. In order to assess goodness-of-fit as explained above the idea of what it means to say that a model fits. Suppose we denote the observed sample values of the outcome variable in fitted value as y where $\hat{y} = (\hat{y}_1, \hat{y}_2, \ldots, \hat{y}_n)$, we conclude that the model fits if; (a) the summary measures the distance between $y$ and $\hat{y}$ are small. (b) the contribution of each pair $(y_i, \hat{y}_i)$, $i = 1,2,3,\ldots,n$, to these summary measures is unsystematic and is small relative to the error structure of the model.

### 4. Model Evaluation Metrics

Mean Squared Error (MSE)

The most common metric for regression is MSE. An estimator's mean squared error (MSE) measures the average of the squares of the errors or the average squared difference between the estimated values and the actual values. The MSE is a measure of the quality of an estimator – it is always non-negative, and values closer to zero are better. The goal is to minimize this mean, which will provide us with the best line that goes through all the points.

$$MSE = \frac{1}{n}\sum_{i=1}^{n}(Y_i - \hat{Y}_i)^2 \quad (14)$$

Where, n = number of data observations, $Y_i$ = observed values, $\hat{Y}_i$ = predicted values.

R-Squared

In a regression model, R-squared (the coefficient of determination) is a statistical measure that quantifies how much of the variance in the dependent variable can be explained by the independent variables. The R-Squared term, displays how well the data fit the regression model (the goodness-of-fit). However, this measure should not solely be the measure when evaluating the goodness of a statistical model, even though it provides some insightful information about the regression model and that a high r-squared value does always imply a good regression model. Nevertheless, A low r-squared value is often regarded as a bad indication for the prediction model.

$$R^2 = 1 - \frac{\sum_{i=1}^{n}(Y_i - \hat{Y}_i)^2}{\sum_{i=1}^{n}(Y_i - \bar{Y})^2} = 1 - \frac{SSR}{SST}$$

(15)

Where, SSR is defined as the sum of squares regression (explained sum of squares), SST is the total sum of squares. However, the effectiveness of the model in fitting the observed data is determined by the SSR.

## IV. RESULT AND DISCUSSION

This section of the paper provides an empirical analysis of the data collected and furthermore discusses the findings of this paper. This section also provides an analysis of the yearly mortality rate and the factors that affect the longevity of SSNIT pensioners in Ghana with data obtained from the Social Security and National Insurance Trust.

### A. Demographics

As shown in Table 1, the total number of male deaths was significantly greater than the total number of female deaths in all years under study. In other words, the number of male pensioners who died between 2015 and 2020 is about four times more than the number of female pensioners who died within the same year scope. However, the mortality rate of the pensioners decreased from 1.24 to 0.53 per 1000 pensioners. There was sufficient evidence that there exists a lower rate of death among female pensioners as compared to male pensioners.

Table 1. The Descriptive information of the SSNIT Pensioners.

| Year | Number Of Deaths | Number Of Pensioners | Mortality Rate per 1000 | Number Of Survivors | Proportion | Number Of Female Deaths | Number Of Male Deaths |
|---|---|---|---|---|---|---|---|
| 2015 | 1942 | 156262 | 1.24 | 154320 | 0.98757 | 306 | 1636 |
| 2016 | 2392 | 174146 | 1.37 | 171772 | 0.98636 | 400 | 1992 |
| 2017 | 2485 | 189549 | 1.31 | 1992 | 0.98688 | 479 | 2006 |
| 2018 | 2542 | 200000 | 1.27 | 197458 | 0.98729 | 470 | 2072 |
| 2019 | 2311 | 215850 | 1.07 | 213539 | 0.98929 | 463 | 1848 |
| 2020 | 1196 | 227407 | 0.53 | 226211 | 0.99474 | 290 | 906 |
| Total | 12868 | 1163214 | 1.11 | 1150364 | | 2408 | 10460 |

*Source: Researcher's field data (2022)*

### 1. Mortality Rates

The mortality rate as explained earlier in this section is the proportion of the number of people who died within a particular cohort by the total number of people within that cohort at a given time. It is observed in Table 2 that, over the five years, the female pensioners' mortality rate is relatively lower than that of the male pensioners. Comparing the mortality rates, it is therefore sufficient to say that there is a lower rate of death among female pensioners as compared to male pensioners. Furthermore, the mortality rate among pensioners decreased with increasing time, i.e., the mortality rate per 1000 pensioners in 2015 was 1.24, however, this value decreased to 0.53 in 2020 as shown in Figure 1 below.

Table 2. The Rate of Mortality for Males and Female SSNIT Pensioners.

| Year | Male Pensioner Mortality Rate | Female Pensioner Mortality Rate | Mortality Rate per 1000 pensioners |
|---|---|---|---|
| 2015 | 0.842430484 | 0.157569516 | 1.24 |
| 2016 | 0.83277592 | 0.16722408 | 1.37 |
| 2017 | 0.807243461 | 0.192756539 | 1.31 |
| 2018 | 0.815106216 | 0.184893784 | 1.27 |
| 2019 | 0.79965383 | 0.20034617 | 1.07 |
| 2020 | 0.757525084 | 0.242474916 | 0.53 |

*Source: Researcher's field data (2022)*

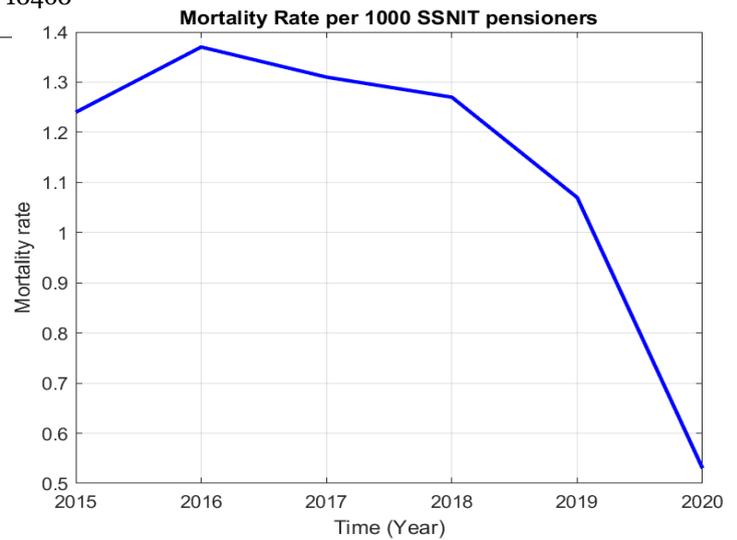

Figure 1. Mortality Rate per 1000 SSNIT pensioners.

### 2. Regression Analysis

**Linear Regression**

Some results from the regression analysis of the data collected is presented in this subsection of the paper. The significance of the variable of interest was assessed and the overall adequacy of the models was examined and presented. Furthermore, the findings of the model fitting, model selection and prediction accuracy of the regression models evaluated using the logistic classifier of the Supervised Machine learning algorithm are presented.

As shown in Table 3, the linear regression model of the longevity of SSNIT Pensioners in Ghana is presented. It can be observed that factors such as basic salary, number of years of service, gender, and ages of the pensioners with p-values of 0.000 each, significantly affected the longevity of the SSNIT pensioners. Furthermore, the model significantly explained 55% of the variability in the model.

The fitted multiple linear regression model is given as

$$y = -29.53 + 0.0000682 \text{basic salary} - 0.4597 \text{service years} - 0.5759 \text{gender} + 0.8115 \text{age}$$

Where y = the number of years survived after retirement (longevity of pensioner), basic salary = basic salary of pensioner, service years = the number of years of service in the workforce, gender = the gender of the pensioner, age = age of pensioner before death.

Table 3. Parameter estimates for the multiple linear regression model.

|  | Intercept | Basic Salary | Number of years of service | Gender | Age |
|---|---|---|---|---|---|
| Coefficients | -29.528 | -6.8217e-05 | -0.45973 | -0.57593 | 0.81147 |
| SE | 0.57388 | 5.8167e-06 | 0.0052813 | 0.08144 | 0.008802 |
| p-value | 0.000 | 0.000 | 0.000 | 0.000 | 0.000 |
|  |  |  | R-squared | 0.5468 | |
|  |  |  | Adjusted R-Squared | 0.5466 | |
|  |  |  | RMSE | 3.6 | |

*Source: Researcher's field data (2022)*

Furthermore, the author juxtaposed the variability of the data using different regressors specifically, the linear regression, decision tree regression, random forest regression and the support vector regression models. It can be observed from Table 4 that the linear regression model had a mean square error and an R square value of 12.17360256 and 0.55667945 respectively. Furthermore, the random forest regression similarly had a mean square error of 12.42185797 and an R squared value of 0.54763884 which explained 55 percent of the variability in the model. Lastly, the decision tree regression and the support vector regression had a mean square error and R-squared value of ($MSE = 20.03918727$, $R^2 = 0.27024202$ and $MSE = 25.27587267$, $R^2 = 0.07954002$) respectively.

Table 4. Comparison of different regression models on longevity of SSNIT pensioners.

| Regressor | Mean Squared Error | R Squared Score |
|---|---|---|
| Linear Regression | 12.17360256 | 0.55667945 |
| Decision Tree Regression | 20.03918727 | 0.27024202 |
| Random Forest Regression | 12.42185797 | 0.54763884 |
| Support Vector Regression | 25.27587267 | 0.07954002 |

*Source: Researcher's field data (2022)*

**Logistic classification model.**

In this section, the author sought to construct a predictive logistic model which would predict the longevity of a SSNIT pensioner provided the pensioner had survived less or more than a certain age specifically, 8 years after retirement. The longevity of the pensioner was then categorized with a value of 1 if the pensioner had survived less than 8 years after retirement and 0 if the pensioner had survived 8 or more years after retirement. As shown in Figure 3, a significant number of the SSNIT pensioners used in the analysis survived less than 8 years after retirement before their death. More specifically, 8000 out of the 12,868 pensioners survived less than 8 years after retirement.

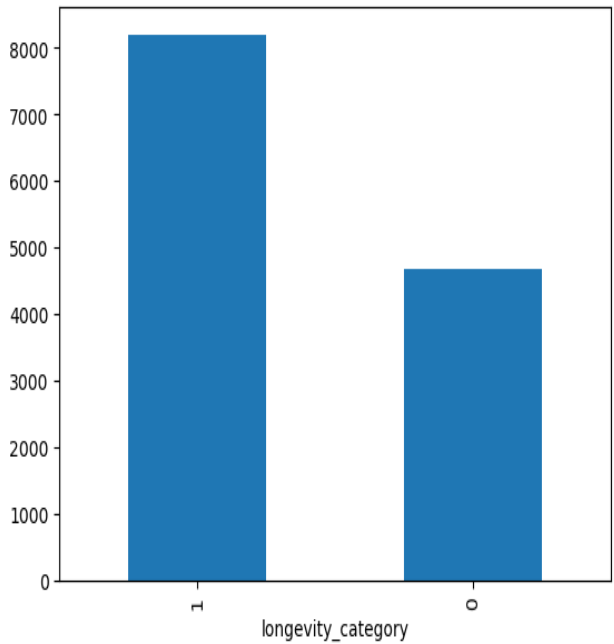

Figure 2. SSNIT pensioners who lived less or more than 8 years after retirement (1 = less than 8 years, 0 = 8 years and above).
*Source: Researcher's field data (2022)*

The author assessed the averages of factors such as the basic salary of pensioners, number of years of service, ages and the number of years survived after retirement (longevity of pensioner) for both categories i.e., pensioners that survived less than 8 years after retirement and pensioners who survived 8 years and above after retirement. As depicted in Table 4, the average basic salary of the pensioners who survived less than 8 years after retirement was GH₵ 7741.827 and served for 35.65 years on average. On the contrary, it can be observed that pensioners who survived 8 years or more after retirement had an average basic salary of GH₵ 5544.20 and served for 31.49 years on average which in juxtaposition is less than that of the pensioners who had survived less than 8 years after retirement. On their age distribution, SSNIT pensioners who survived less than 8 years had a mean age of 63.51 years and pensioners who lived 8 years and above after retirement had a mean age of 67.29 years. This shows that pensioners who survived 8 years and above after retirement spent relatively few numbers of years serving in their various workforce and survived longer than pensioners who served longer in the workforce but survived less than 8 years after their retirement.

Table 5. Mean of variables of interest of longevity of pensioners less or more than 8 years.

| Longevity_category | Basic Salary | Number of years of service | Age | Longevity of Pensioner |
|---|---|---|---|---|
| 0 (8 years and above) | 5244.198322 | 31.491758 | 67.294798 | 12.343395 |
| 1(less than 8 years) | 7741.826979 | 35.652537 | 63.509313 | 3.681653 |

*Source: Researcher's field data (2022)*

As shown in Table 6, is the summary of the logistic regression model of the determinants of longevity (less than 8 years or 8 years and above after retirement before death) of SSNIT pensioners in Ghana. The intercept which indicates the baseline effect on the number of years survived after retirement before death (longevity of pensioner), has an estimate of 28.65 with an odds ratio of $2.775 \times 10^{12}$, was statistically significant at a 5% significance level. The predictor basic salary with an odds ratio of 1.0000011 and a p-value of 0.000 indicated that, with all the other predictors held constant, an increase in the basic salary of the pensioner significantly increases the odds of the pensioner surviving less than 8 years after retirement by 0.0011%. A unit increase in the number of years of service (OR = 1.2236; p-value = 0.000), significantly increases the likelihood of a SSNIT pensioner surviving less than 8 years after retirement by 22.36% with all other factors held constant at a 5% significance level. Moreover, with all other predictors held constant, the gender (OR=1.4695; p-value = 0.000) significantly increases the likelihood of a SSNIT pensioner surviving less than 8 years after retirement by 46.95% a 5% significance level. Furthermore, the log odds ratio between the given level (Male) and the reference level (Female) is given as 0.3849. Lastly, for the age of the pensioner before death (OR = 0.5801; p-value = 0.000), significantly at 5% reduces the odds of a SSNIT pensioner surviving less than 8 years after retirement by 41.98%.

Predictors such as basic salary, the number of years of service in the workforce, the age of pensioner before death and the gender of the pensioner with p-values of 0.000 each were statistically significant in predicting the number of years a SSNIT pensioner survived after retirement before death that is, whether a particular SSNIT pensioner survived less or more than 8 years after retirement.

The logistic model is given as

$$In\left(\frac{\pi}{1-\pi}\right) = 28.65 + 0.000001076\lambda_1 + 0.2019\lambda_2 + 0.3849\lambda_3 - 0.5445\lambda_4$$

Where $\lambda_1$ = basic salary of pensioner, $\lambda_2$ = the number of years of service in the workforce, $\lambda_3$ = the gender of the pensioner, $\lambda_4$ = age of pensioner before death.

Alternatively,

The fitted Model is given as

$$\pi = \frac{e^{(28.65+1.076e-06\lambda_1+0.2019\lambda_2+0.3849\lambda_3-0.5445\lambda_4)}}{1+e^{(28.65+1.076e-06\lambda_1+0.2019\lambda_2+0.3849\lambda_3-0.5445\lambda_4)}}$$

(1)

Where $\lambda_1$ = basic salary of pensioner, $\lambda_2$ = the number of years of service in the workforce, $\lambda_3$ = the gender of the pensioner, $\lambda_4$ = age of pensioner before death.

Table 6. Parameter estimates of the logistic regression model for longevity.

|  | Intercept | Basic Salary | No of years of service | Gender | Age |
|---|---|---|---|---|---|
| Coefficients | 28.65 | 1.076e-06 | 2.019e-01 | 3.849e-01 | -5.445e-01 |
| SE | 6.027e-01 | 7.415e-06 | 5.054e-03 | 6.719e-02 | 1.035e-02 |
| Wald Test | 47.53609 | 0.1451113 | 39.94856 | 5.728531 | -52.6087 |
| p-value | 0.0000 | 0.0000 | 0.0000 | 0.0000 | 0.0000 |
| Odds ratio | 2.7752e12 | 1.0000108 | 1.223683 | 1.469467 | 5.801037e-01 |

Source: Researcher's field data (2022)

Next, we analyze the accuracy of this model by constructing a confusion matrix as shown in Table 7 below.

As presented in Table 8 below, Metrics such as accuracy and balanced accuracy were used in the classification report to assess a model's performance across various classes. That is, classifying the number of years pensioners survived, whether less than 8 years (1) or 8 years and above (0) after retirement using predictors such as basic salary, age, gender, and the number of years of service. The precision of the model's positive predictions is a measure of their accuracy. Out of all instances predicted as belonging to a class, these values show the percentage of correctly predicted instances for each class. As depicted in Table 7 below, there were 3418 instances for class 0, where the logistic model correctly classified a SSNIT pensioner who had lived less than 8 years after retirement to indeed have survived less than 8 years after retirement and 7270 instances for class 1, where the logistic model correctly classified a SSNIT pensioner who had lived 8 or more years after retirement to indeed had lived 8 or more years after retirement respectively. Moreover, an accuracy score of 0.83 as shown in Table 8, indicates that the model 83% of the time accurately predicted the class labels that is, whether a pensioner had lived less than 8 years (1) or above 8 years (0) after retirement with a 95% Confidence Interval (82.4%, 83.7%) which depicts a very good model.

Table 7. Confusion Matrix for Logistic Regression Model.

|  | 0 (8 years and above) | 1 (less than 8 years) |
|---|---|---|
| 0 (8 years and above) | 3418 | 927 |
| 1 (less than 8 years) | 1253 | 7270 |

*Source: Researcher's field data (2022)*

Table 8. The Classification Report for model.

| | |
|---|---|
| **Accuracy** | **0.8306** |
| **95% CI** | (0.824,0.837) |
| **Balanced Accuracy** | 0.8093 |
| **Mcnemar's Test P-Value** | 3.385e-12 |
| **'Positive' Class** | 0 |

*Source: Researcher's field data (2022)*

## V. CONCLUSION

In conclusion, this paper sought to determine the significant determinants of longevity amongst pensioners in Ghana, more specifically SSNIT pensioners by estimating the mortality rate of SSNIT pensioners in Ghana, determining the factors that significantly affect the longevity of the average Pensioner in Ghana and constructing a predictive model for predicting the longevity rate among SSNIT pensioners in Ghana. This paper utilized secondary data which was obtained from the leading pension provider in Ghana, the Social Security and National Insurance Trust, which is mandated to manage the First-tier pension scheme in Ghana. The data spanning from the periods of 2015 through 2020 was used for the development of the regression models. This dataset consisted of variables such as gender, year of entry, year of retirement, basic salary, date of birth and date of death of pensioners amounting to 12,868 data points. The paper utilized models such as multiple linear regression, Decision Tree Regression, Random Forest Regression, Support Vector Regression, and binary logistic regression at different sections. The findings of this paper revealed that there exists a lower rate of death among female pensioners as compared to the male pensioners that is, over the five years, the female pensioner mortality rate was relatively lower than that of the male pensioners about four times greater.

Furthermore, the study revealed that factors such as the basic salary of the pensioner, the number of years of service a pensioner spent in the workforce, gender and age of the pensioner were significant determinants for determining the number of years a pensioner survived after attaining the retirement age of 60 years at a significance level of 5%. The model was able to explain approximately 55% of the variability in the model. Juxtaposing the variability of the data using different regressors specifically, the linear regression, decision tree regression, random forest regression and support vector regression models, the paper indicated that the linear regression model better explained the variability in the data than the other regression models. Moreover, to be able to construct a predictive logistic model which would predict the longevity of a SSNIT pensioner provided the pensioner had survived less than a particular age or the particular age and above specifically, 8 years after retirement. The longevity of the pensioner was then categorized with a value of 1 if the pensioner had survived less than 8 years after retirement and 0 if the pensioner had survived 8 or more years after retirement before death. Further analysis revealed that the average basic salary of the pensioners who survived less than 8 years after retirement was GH₵ 7741.827 and served for 35.65 years on average. On the contrary, it can be observed that pensioners who survived 8 years and above after retirement had an average basic salary of GH₵ 5544.20 and served for 31.49 years on average.

Lastly, the fitted binary logistic regression model constructed had an accuracy score of 0.83 indicating that the model 83% of the time accurately predicted the class labels that is, whether a pensioner had lived less than 8 years (1) or above 8 years (0) after retirement. The findings of this paper revealed that determinants such as basic salary, the number of years of service in the workforce, the age of the pensioner before death and the gender of the pensioner were statistically significant in predicting the number of years a SSNIT pensioner survived after retirement before death that is, whether a particular SSNIT pensioner survived less or more than 8 years after retirement. The authors recommend to Ghanaian employees that, their health should be of a great priority as the number of years of service in the workforce keenly determines the number of years they would survive after retirement before death as every increase in the service years increases the likelihood of a pensioner surviving less than 8 years after retirement by 22.36%. This paper utilized solely GLMs in assessing the factors that affect the longevity of SSNIT pensioners in Ghana. In the Ghanaian scene, there are other emerging social security schemes aside from the well-known one provided by SSNIT, further comparative studies can be conducted to investigate which factors affect the longevity rate of pensioners on these social security plans as compared with SSNIT pensioners. Furthermore, other studies can investigate the longevity rate and further forecast the longevity rate of Ghanaian Pensioners using more sophisticated stochastic models such as the Lee-Carter Model (Lee & Carter, 1992) to inform decisions by policymakers, government and insurers as well as employees in Ghana.

## VI. ACKNOWLEDGEMENT

The author would like to acknowledge the Social Security and National Insurance Trust (SSNIT) which is the leading pension provider in Ghana across all sectors of the Ghanaian economy for the availability of the secondary data on SSNIT pensioners. Furthermore, the author appreciates all contributors and reviewers for their insightful contributions.